# AN EXAMINATION OF SKILL REQUIREMENTS FOR AUGMENTED REALITY AND VIRTUAL REALITY JOB ADVERTISEMENTS

*Amit Verma\*, Craig School of Business, Missouri Western State University, Saint Joseph, MO, USA*
*Pratibha Purohit, School of Business, University of Connecticut, Storrs, CT, USA*
*Timothy Thornton, College of Education, Athens State University, Athens, AL, USA*
*Kamal Lamsal, Craig School of Business, Missouri Western State University, Saint Joseph, MO, USA*

*\*Corresponding Author: averma@missouriwestern.edu, Address: 315L, Popplewell Hall, 4525 Downs Drive, Saint Joseph, MO, 64507, Tel: +1 816 271 4357, Fax: +1 816 271 4338*

**ABSTRACT**

*The field of Augmented Reality (AR) and Virtual Reality (VR) has seen massive growth in recent years. Numerous degree programs have started to redesign their curricula to meet the high market demand of such job positions. In this paper, we performed a content analysis of online job postings hosted on Indeed.com and provided a skill classification framework for AR/VR job positions. Furthermore, we present a ranking of the relevant skills for the job position. Overall, we noticed that technical skills like UI/UX design, software design, asset design and graphics rendering are highly desirable for AR/VR positions. Our findings regarding prominent skill categories could be beneficial for the human resource departments as well as enhancing existing course curricula to tailor to the high market demand.*

**INTRODUCTION**

According to the Hired.com survey (Hired, 2021), Augmented Reality (AR) and Virtual Reality (VR) interview requests have grown by 1400% over the last year. These jobs have been identified among the top emerging jobs of the year (Perry, 2020). 74% of those surveyed by Hired.com agreed that the technology would have a major impact on the job market within five years. The demand for these job positions has grown 79% compared to last year, partly due to rapid advancements in AR/VR inventions and patents. It has been forecasted that by 2030 over 23 million jobs will be impacted by the two technologies (Alsop, 2021). The same findings were echoed by a Facebook study (Facebook, 2021) which estimated substantial adoption of AR/VR across all industries. Several industries have already begun incorporating AR/VR technologies to enhance customer experiences. For example, Ikea, Lacoste, New York Times, Kate Spade, etc. have leveraged the technology to aid the marketing delivery channels (Paine, 2021). Both studies noted that the market demand for AR/VR jobs is prominent and growing. Even though there is a forecasted increase in demand, there remains a dearth of skilled professionals. Similar skill shortages exist in other newer Information Technology (IT) domains as well as traditional fields like marketing, finance, healthcare, and supply chain (Wilson et al., 2017). The AR/VR job opportunities are dispersed across various industries. Because of the evolving nature of work and the need for standardization, better classification of the job skills in this domain will make the hiring process more coordinated and efficient. Moreover, the structure provided by such a classification could be utilized in curriculum design to address the skill gaps and highlight the relatively important skillsets. Therefore, one of the goals of this research project is to develop a comprehensive skill classification framework for AR/VR job positions.

Content analysis was utilized to extract keywords from job postings hosted on Indeed.com in order to calculate the frequencies of specific skills corresponding to each skill category. In this way, we identified the most critical skills in the dataset which was based on the relative frequency of each skill for AR/VR jobs. The findings can also aid in curriculum design efforts such that the most valued skill sets are emphasized in the existing analytics curricula. Moreover, the existing AR/VR courses can be revised to reflect the current sought-after software tools and applications.

The studied job positions belong to the professional areas of AR and VR. Both of these domains were defined within the big umbrella term of Mixed Reality (MR) or Extended Reality (XR) based on Billinghurst et al.(2017). MR merges real and virtual worlds to create new environments. The relationship between these fields is summarized by the mapping in Figure 1.

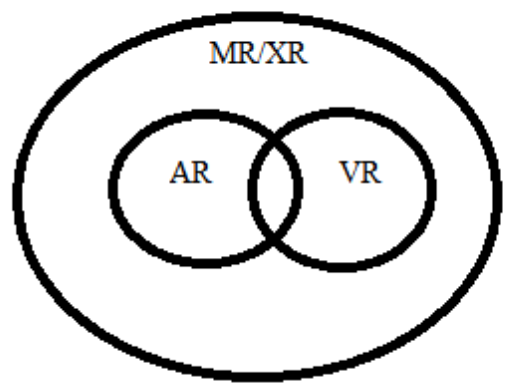

**Figure 1**. Hierarchy Of Different Domains

The three areas immerse a digital layer (typically through a smartphone or headset) over the physical reality. The individual differences are due to varying levels of immersion and the relationship between the two layers. AR enhances an individual's physical environment by overlaying an interactive 3D virtual object on a real-world environment (Thornton et al., 2012). AR supplements the physical reality by adding only some extra details. The classic examples of AR include Google Glass, Pokemon Go, and Snapchat filters. On the other hand, VR incorporates the highest level of immersion and fully immerses the user in a computer created environment (Goldiez et al., 2004). VR users typically wear a headset like Oculus Rift and Microsoft Holoens to engage in a completely virtual experience. Note that some simple adaptations of VR have been initiated by Google Cardboard using only a smartphone and cardboard. MR or XR allows physical and digital layers to operate in tandem and might involve different combinations of AR and VR. Gaming consoles such as Xbox Kinect rely on Human Computer Interaction (HCI) coupled with gesture recognition and sensors to operate in this domain. The higher-end applications of MR also exist in drone and robot-assisted surgeries.

The interest in AR/VR technologies has especially grown since the pandemic (Robertson, 2021; Facebook, 2021). Academic institutions could benefit from an in-depth exploration of the skill requirements in this field. In this regard, the goal of this study was to address the following research questions:

1. Which skills are required for AR/VR professionals?
2. What majors do employers request for AR/VR professionals?
3. What software tools are currently in-demand for such positions?

In this paper, job advertisements in the US were studied. A breakdown of different skills required for the AR/VR positions was analyzed in order to provide insights into employers' expectations. Using the ranked skill categories based on relative frequency, academics can repurpose the existing degree programs in order to highlight the current skillsets required in the job market. The focus of this study was the "demand side" of the US job market via a descriptive analysis of the job advertisements posted by employers. Future research could analyze the skills gaps by studying the "supply side" of the related degree programs. This would involve a review of the course descriptions or course syllabi. Scholars could also perform a survey of administrators and faculty members who are responsible for course design in the AR/VR domain. Another future research direction involves the study of local trends in the job market across multiple countries or multiple regions within the same country.

The main contributions of this study were:

1. The identification of the key skills needed for AR/VR based on our skill classification framework.

2. The analysis could assist the human resources team in creating job advertisements in addition to specific current training modules for the working professionals in the industry.
3. Academic institutions can also utilize our findings in curricula design efforts. This is helpful to create new courses or restructure existing course offerings. Moreover, the students interested in related career domains could also emphasize the top in-demand skills in their job applications.

This paper contributes to the extant literature of curriculum design in degree programs by presenting the popular skills in the AR/VR domain. Thus, universities could align their coursework with the existing job market demand. In addition, employers could design more standardized and better-curated job positions, clearly emphasizing the required skills.

The article was organized as follows. Initially, a literature review on the impact of AR/VR on various disciplines was presented, followed by the employers' perspective of job requirements. Next, the classification framework, research technique, and datasets used for the study were discussed. Thereafter, the results detailing a ranked list of the skill categories sorted with respect to the relative frequency among the total number of jobs was presented. Finally, the conclusions of the study and an outline of the future research directions was discussed.

## LITERATURE REVIEW

AR/VR is an emerging technology with wide-ranging applications in many domains. The technology augments the physical space with a digital presence in order to enhance customer experiences (Schmalstieg & Höllerer, 2016). The nascent phases of this technology found applications in video games. However, over the last decade, various industrial and commercial applications have been envisioned. Various applications of AR/VR have been proposed for robotics, engineering, automotive, aerospace, maintenance and education (Nee at al., 2012; Chi et al., 2013; Akçayır & Akçayır, 2017; Doshi, 2017; Mourtzis et al., 2017; Helin et al., 2017). The standard definitions and related taxonomies, along with some successful applications, are surveyed in Billinghurst et al. (2017). The impact of AR/VR on the field of education has been studied by the authors in Bacca Acosta et al. (2014). The different pedagogical approaches involved in teaching and training along with case studies is detailed in Wang et al. (2018). In terms of STEM learning, the review of the literature of AR/VR is provided in Ibáñez and Delgado-Kloos (2018). Similar to the technique used in this study, a content analysis was utilized through a study of 28 research publications from 2010 to 2017. Finally, the study presented AR/VR activities that could be used a part of instructional delivery.

A set of closely related studies were proposed by Fomiykh et al. (2019), Fomiykh et al. (2020a) and Fomiykh et al. (2020b). First, Fomiykh et al. (2019) analyzed the existing course offerings in AR/VR as well as conducted a preliminary job market analysis. The study summarized the various teaching methods, learning activities, course objectives and learning outcomes. The authors compiled a list of skill sets based on the firsthand scan of few job postings followed by a survey of industry professionals. More specifically, they studied only the technical skill requirements for 16 jobs in the US and the coverage of the study was global. Given that the biggest players in the AR/VR domain viz. Apple, Facebook and Google belong to the US, an in-depth assessment of the skill requirements is warranted. Hence, 397 job advertisements in the US using a complete skill classification framework consisting of hard and soft skills were analyzed in this paper. Fomiykh et al. (2020a) and Fomiykh et al. (2020b) provided a blueprint for the CS curriculum in order to facilitate AR/VR teaching in universities. They also provided an outline of two AR/VR courses at the foundation and advanced levels using 12 groups of skills. The study is concluded with a discussion on curriculum design implications.

AR/VR can be considered as one of the digital skills needed for success in the workplace of the future. The technical and soft skill sets needed for Industry 4.0 is analyzed in Saari et al. (2021). In addition to Industry 4.0, other relevant fields are business analytics, data analytics, and big data. Many studies have explored the employers' side through the skill-based analysis on job postings. For instance, Debortoli et al. (2014) analyzed the difference between big data and business intelligence job positions using a skill classification framework based on a Latent Semantic Analysis of the job descriptions. The authors concluded that the business domain skills as well as technical skills were valued for both job positions. Gardiner et al. (2018) studied 1216 job advertisements having "big data" as part of their title. The authors found that analytical skills and soft skills are highly valued for such job positions in addition to technical skills.  Lovaglio et al. (2018) collected Information and Communication Technology (ICT) and Statistical Italian job advertisements between June and September 2005. The authors concluded that statistical positions require more technical and computing skills in relation to soft skills. Verma et al. (2019) utilized a skill classification framework for related job categories like Business Analyst, Business Intelligence Analyst, Data Analyst and Data Scientist using a content analysis of 1235 job advertisements. The results indicated that skills such as decision making, organization, communication, and data management were essential for success in these professions. More recently, Anton et al. (2020) analyzed the skill requirements for three related occupations: Data Science and Engineering, Software Engineering and Development, and Business Development and Sales. A mixed-methods approach was employed in conjunction with a text mining of scientific literature. Similar to other studies, the authors presented the important soft and hard skills required for these positions.

The current study supplements the extant literature by proposing a new skill classification framework for AR/VR jobs. Given the high demand for AI/VR jobs, an in-depth analysis of the required skills is merited. Therefore, this paper's goal was to study an important job category in the Industry 4.0 era.

## RESEARCH METHODOLOGY

**Skill Classification Framework**

In this section, the skill classification framework used for AR/VR job postings was presented. The goal was to determine if each job posting was a member of one or more of these skill categories based on the keywords. A list of skill categories developed by Fomiykh et al. (2019), Verma et al. (2021a) and Verma et al. (2021b) was used as a starting point. Additionally, more skill categories to reflect the needs in the AR/VR domain were added. For achieving this, a sample of 50 AR/VR jobs was taken and an in-depth content analysis of the skills needed for AR/VR was conducted. The skill categories and subcategories were then updated based on the findings. Three independent raters were utilized to determine the reliability score of the proposed categorization framework. The alpha coefficient of the intercoder reliability measure (based on De Swert, 2012) was calculated as 0.91. The final list of skill categories and subcategories, along with sample keywords were detailed (see Table 1).

**Table 1**. Classification Framework

| Skill Category | Skills | Keywords |
|---|---|---|
| Communication | Written | Copywriting, Editing, Blogging, Content Creation, Story-ideation |
| | Verbal | Verbal, Oral, Cold calling |
| | Presentation | Present, Presentation, Report |
| | Generic | Responsible, Determined, Competitive, Witty, Success-oriented |
| Employee Attributes | Motivation | Motivated, Ambition, Willingness to learn, Delivering result, Continuous learning |
| | Time Management | Time management, Timely manner, Prioritize time, Deadline driven |
| | Detail oriented | Attention to detail, Eye for detail, Accuracy, Precision |
| | Attitude | Can do, Go-getter, Self-learner, Self-directed, Positive Attitude |
|  | Independence | Independence, Without supervision, Autonomous |
|  | Adaptability | Adaptable, Flexible, Multitasking |
|  | Confidence | Confident, Decisive |
|  | Other | Funny, Smiling, High energy, Reliable, Proactive |
| Occupational Attributes | Programming | Python, C#, C++, VB, Excel Macros, PERL, C, Java, Visual Basic, VB.NET, VBA, COBOL, FORTRAN, S, SPLUS, BASH, Javascript, ASP.NET, JQUERY, JBOSS |
|  | Software Design Principles | API, REST, Product lifecycle management, JSON, REACT, Architecture, Testing, Security, Source control, Unit testing, Debugging, Documentation, Continuous integration, Build trains, OOP, Event handling, Source control, Inheritance, Abstraction, Encapsulation, Loops, Control-logic, Multithreaded, Data structure, Devops |
|  | XR SDKs | UIKit, AVFoundation, Core Motion, Core ML, CloudKit, SiriKit, StoreKit, ARKit, RealityKit, ARKit, ARCore, AR Foundation, MRTK, WebXR, CoreMedia, CoreAudio, CoreAnimation |
|  | Asset Design | Prototype, Visualization, Visual effect, 3D asset, asset, content developer, content strategist, 3D images, Animation, Proof-of-concept, Storytelling, Validation, Content, Film production, Video production |
|  | UI/UX Design | Interaction, Studying users, XR experience, UX Design, Interaction design principles, Usability, Design thinking process, HCI, Accessibility, User experience design, DFI, DFM, DFA |
|  | 3D Software | Maya, Houdini, Blender, 3Ds Max, Arnold, RenderMan, Cycles, three.js, Meshlab, Blender, Maya and Cinema4D, Direct3D MAYA, 3D MAX, Autodesk 3D, Autodesk 3ds max, BIM 360 |
|  | Game Engine | Unity, Unreal |
|  | XR Hardware | HTC VIVE, Varjo HMD, Microsoft Hololens, Rift, VIVE, Gear VR, Oculus Rift, PlayStation VR, Google Cardboard, Google Daydream, Samsung Gear VR, Magic Leap, Rift, Oculus Quest, Microsoft Hololens |
|  | Performance Tools | VTune, XPerf/GPUView, valgrind, Instruments, Performance profiling, Profiling, Profiler |
|  | 2D Software | Adobe Sketch, Illustrator, FIGMA, CAD, CAM, Adobe Illustrator, Autodesk, Revit, Adobe Photoshop, Photoshop, Adobe Sketch, Illustrator, Metal, FIGMA |
|  | Graphics Rendering | Metal, HLSL, GLSL, Animation effects, VFX, Computer graphics, Imaging system, Imaging, Feature definition, Computational photography, Object capture, Rendering, Visual effect, Image signal processing, Texture, Lighting, Photogrammetry, Modeling, Mapping, Rigging |
|  | Mathematics | Geometry, Linear Algebra, 3D math, 3D geometry, Vector Math, Vector |
|  | Sensors | EEG, fMRI, fNIRS, EMG, MTF, SNR, human interface, HCI, haptic, hand gesture, compression, sensing, PICO |
|  | Cloud Tools | AWS, Google Firebase, Azure, Server |
|  | Project Management | Project management, PERT, CPM, PERT/CPM, Change management, Project budget, Project documentation, PMP, Microsoft Project, Gannt Chart, Lean, Agile |
|  | General Hardware | Hardware, Architecture, Devices, Printer, Storage, Desktop, PC, Server, Workstation, Mainframe, Legacy, System architecture |
|  | Decision Making | Reporting, Analysis, Modeling, Design, Problem-solving, Implementation, Testing, Analytical, Strategic thinking |
| Interpersonal | Interpersonal | Team management, Collaboration, Cooperation, Networking, Client relationship |

| Problem Solving | Problem Solving | Problem solving, Troubleshoot, Conflict resolution, Solve issue, Critical thinker |
|---|---|---|
| | Creativity | Creative, Out of box, Storyteller |
| | Process Design | Design Process, Improve process, Continuous improvement, Operations management |
| Administrative | Administrative | Issue management, Posting schedule, Product launch, social calendar |
| Analytical | Analytical | Insight, Identify trend, Summarize finding, Analyze trend, Synthesize information, Draw conclusion, Propose solution, Google Analytics, ArcGIS, GIS, QGIS, Data Analytics, Business Analytics |
| Research | Research | Data gathering, Data collection, Data reporting, Monitor trend, Monitor performance |
| Numeracy | Numeracy | Financial Management, Bookkeeping, Accountancy |
| Foreign Language | Foreign Language | Spanish, French, German, Italian, Chinese |

**Research Technique**

For this research project, job postings from Indeed.com, the most popular job search site across the US were utilized. Web scraping was the primary source of data collection. The data was extracted during June – December 2020. The results on a web search on Indeed.com based on the titles of "Augmented Reality (AR)" and "Virtual Reality (VR)" were gathered. After downloading the job postings through Python, the job description was split up into keywords using content analysis. Content analysis is a tool to analyze words in a document and identify patterns in the text (Neuendorf, 2016). The text was divided into various categories in order to summarize the skill requirements for the studied job positions. Note that each of these individual keywords of the text fits into a specific category and subcategory given by the skill classification framework described in the previous section. This allowed the authors to measure the required skills for each job posting. In this way, aggregate measures were developed for the relative importance of each skill category (see Figure 2).

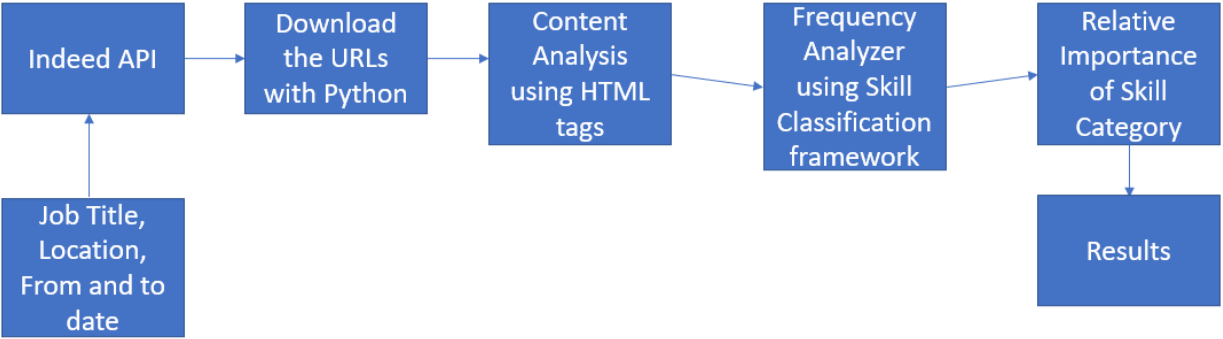

*Figure 2: Summary of our research technique*

Next, each component of the technique was detailed. The first component was the web scraper. Python was utilized to build the web scraping tool. It was entirely dependent on the existing Indeed Application Programming Interface (API). The API requires job title, location, and time period as input arguments. The specific search query based on job titles was conducted during the specific time period in the United States. The output was an XML file consisting of the job title, job URL, location, company, posting date, and a job summary. This job summary field provided a very high-level summary of the job description but did not contain all the necessary information. Thus, the job URL becomes critical for our approach because the job summary field of the API did not contain important information such as software development kits (SDKs), programming languages, graphical design tools, and specific hardware. To address the lack of relevant information, the actual job postings corresponding to the URLs provided by the API were downloaded.

The downloading procedure was also completed programmatically using Python leading to the local storage of HTML files. Each HTML file contained different elements stored under different tags. The benefit of using the Indeed.com API output was that the final webpage followed a consistent design. Therefore, the job descriptions were always found using a specific tag. Thus, the development of the parser could be streamlined. In turn, the parser extracted the job description for each downloaded HTML file. The job description field was crucial since it contains important data points such as job type, major, related work experience, and required software tools. Furthermore, the parser eliminated all unnecessary keywords from consideration like a, an, the, in, for, and special symbols like quotation marks and semicolon. The parser is paramount for success of natural language processing projects involving content analysis (Neuendorf, 2016). The residual words were individually recognized as unigrams. The two-closest and three-closest unigrams were considered as bigrams and trigrams respectively. These n-grams were critical to our technique because the keywords belonging to each skill category (detailed in Table 1) were directly matched with the n-grams.

The parsed text within the job description tag was inputted to a frequency analyzer that counts whether the text contains specific keywords belonging to each category and subcategory of our skill classification framework. Recall that the associated keywords for each skill category were listed in Table 1. If any keywords belonging to a specific skill were present in the job description, the authors declared that the associated skill category was required for a specific job posting. This method was repeated for each job posting and each skill category. Combining all the outcomes, the number of jobs for which a specific skill category was required were calculated. This characterized the frequency of occurrence of each skill category for each job title. In this way, the relative importance of each skill category for AR/VR job title was determined. Note that the relative importance of each skill category was directly provided by the relative frequency (given by the frequency count divided by the total number of jobs). The relative frequency was reported as a percentage count (see Table 4). This data will be employed in the next section to establish the relative importance for different skill categories.

## RESULTS AND DISCUSSION

The job titles containing phrases "Augmented Reality / AR" and "Virtual Reality / VR" between July and December 2020 were obtained from Indeed.com. A large sample size of 397 was the target for data analysis. The top five states wherein the jobs were located was developed (see Table 2).

**Table 2**. Geographical Distribution of AR/VR Jobs

| Rank | State (Percentage Count of AR/VR jobs) |
|---|---|
| 1 | CA (90.20%) |
| 2 | CO (5.33%) |
| 3 | MA (2.10%) |
| 4 | NY (1.13%) |
| 5 | TX (0.32%) |

The majority of the current interest in AR/VR technologies resides in the IT sector. As evident from the results, most of these jobs are located in the Silicon Valley in the state of California. Denver, Colorado has also emerged as an upcoming entrepreneurial hub for newer technologies. IT giants like Facebook, Apple, Google and Amazon have propelled interest in AR/VR technologies, especially in the metropolitan areas of San Francisco.

The required majors for the job positions were systematically extracted using regular expressions in content analysis, which allows matching of strings like Graphic Design in a specific neighborhood of the complete text of job description field. To accomplish this task, complete sentences in the job description of each job posting were used. More specifically, keywords in the neighborhood of information that pertains to majors like B.S., B.A, Bachelor's, etc. were examined. These matched keywords belong to related majors in the AR/VR field, such as Computer Science, Fine Arts, Information Technology, etc. (see Table 3).

**Table 3**. Majors Required for AR/VR Jobs

| Rank | Major (Percentage Count of AR/VR jobs) |
|---|---|
| 1 | Engineering (65.48%) |
| 2 | Computer Science (39.09%) |
| 3 | Machine Learning (27.41%) |
| 4 | Mathematics (21.32%) |
| 5 | Architecture (15.23%) |

As can be observed in Table 3, the percentage column for each skill category does not sum up to 100% because each job posting might accept multiple majors like Bachelor's in Computer Science or Machine Learning. It was clear that the various engineering majors were more desirable for AR/VR professions, given the technical nature of the job. The conventional degree programs in mathematics and computer science were still valued for AR/VR jobs.

Next, a breakdown of the skill requirements for the AR/VR job positions was provided. The five most common skill categories required for the job position were sorted from high to low in terms of the number of job advertisements in which at least one skill associated with the skill category is present (see Table 4). The top five skills associated with each skill category were also presented. These skills were similarly ranked with respect to the relative frequency (expressed as percentage count). For this purpose, the number of job positions associated with a specific skill was divided by the total number of job postings.

**Table 4**. AR/VR Results

| Skill Category | Skill | Percentage Count (%) |
|---|---|---|
| Occupation | | |
| | UI/UX Design | 95.48 |
| | Software Design Principles | 90.36 |
| | Asset Design | 71.07 |
| | Graphics Rendering | 64.47 |
| | Programming | 57.87 |
| Employee | | |
| | Time Management | 64.41 |
| | Motivation | 31.91 |
| | Attention to Detail | 26.40 |
| | Independence | 22.84 |
| | Other | 11.42 |
| Communication | | |
| | General | 60.91 |
| | Verbal | 42.61 |
| | Written | 24.87 |
| | Presentation | 12.69 |
| Interpersonal | | |

|  | Team Management | 50.76 |
|  | Personal | 29.95 |
|  | Creativity | 24.87 |
|  | Other | 15.74 |
| Analytical |  | 39.09 |
| Administrative |  | 21.83 |
| Research Skills |  | 18.27 |

The research project established 38 skills structured into 10 different skill categories for AR/VR professions (see Table 1). While the literature provided some insights regarding AR/VR professions (Fominykh et al., 2019, Fominykh et al., 2020a and Fominykh et al., 2020b), a thorough analysis of hard and soft skills centered on extensive empirical data was conducted by the authors. The occupational skills were assigned the highest priority for AR/VR job postings. UI/UX design skill subcategory was ranked at the top. This included designing user experiences while keeping in mind the constructs of accessibility and human computer interaction (HCI). In this way, people with varying abilities should be able to use the end product seamlessly. The employees were also expected to be familiar with various design concepts like New Product Introduction (NPI), Design for Manufacturing (DFM) and Design for Assembly (DFA). These concepts typically exist in engineering activities and strengthen the connection of AR/VR with Computer Aided Design (CAD) and Computer Aided Manufacturing (CAM). This also emphasized the overwhelming need for majors with an engineering background in these job postings, as illustrated in Table 3. The next ranked subcategory was the traditional software design principles. Thus, successful employees have a concrete understanding of software engineering principles like object-oriented programming, product lifecycle management and security protocols. Employees should also be able to perform traditional software developer tasks, including debugging, testing, documentation, and source control. Thus, they should be proficient in DevOps, which involves the combination of software development activities with business operations. Asset Design is the next prioritized skill subcategory dealing with video production or simulation in order to develop a prototype or proof of concept to meet client requirements. These activities typically involve content creation through storytelling aided via VFX animations or some other 3D modeling software. The list of 3D modeling software is also captured separately through a skill subcategory defined in Table 1. The relevant results in the form of a sorted list based on the percentage occurrence of the 3D modeling tools were presented (see Table 5).

**Table 5**. 3D modeling software for AR/VR Jobs

| Rank | Software (Percentage Count of AR/VR jobs) |
|---|---|
| 1 | Autodesk Maya (23.26%) |
| 2 | CATIA (22.93%) |
| 3 | Autodesk VRED (20.41%) |
| 4 | Blender (9.30%) |
| 5 | Autodesk ReCap (4.65%) |

These findings established that the marketplace was dominated by commercial software from Autodesk (see Table 5). The only open-source offering consists of Blender at the fourth place. These in-demand tools could be utilized at the course level to enhance student learning experiences. In this way, we can well prepare the students for success in the AR/VR industry. The next critical skill category consists of Graphic Rendering. This exercise involves tracking, lighting, rigging and mapping objects in scenes through the use of algorithms relying on deep learning and reinforcement learning. At a fundamental level, this skill assumes familiarity with visual effects using features and spatial imaging. This is sometimes achieved by using a shader programming language to adjust the levels of light, darkness and color. Both High Level Shading Language (HLSL) and OpenGL Shading Language (GLSL) are written in a C-like language. The AR/VR job positions are technical in nature and place a higher emphasis on computer programming skills. The top five programming languages in order of relative occurrence in the job advertisements were identified (see Table 6).

**Table 6**. Programming languages for AR/VR Jobs

| Rank | Programming Language (Percentage Count of AR/VR jobs) |
|---|---|
| 1 | C/C++/C#/Objective C (32.72%) |
| 2 | Java (10.49%) |
| 3 | Python (8.95%) |
| 4 | SQL (5.25%) |
| 5 | Swift (4.63%) |

These findings identified C-like languages as most common. Given that the popular game engines Unity and Unreal are built on the C-type languages, this observation was trivial. Moreover, Android and iOS application development requires knowledge of Java and Swift programming language respectively. Python is needed for deep learning and reinforcement learning for advanced imaging applications in the graphical rendering pipeline. Basic data management skills using SQL are also in-demand.

AR/VR development is facilitated by SDKs designed by Apple, Google and Facebook. Apple utilizes popular frameworks like ARKit and RealityKit, while Google's service is called ARCore. Facebook's immersive experiences are created through AR Studio. Traditional AR/VR job positions also entail game development on web, desktop or mobile environments. The web-based framework for AR/VR is called WebVR or WebXR, a Javascript application that allows interaction with various AR/VR devices. The ranked list of relative frequency of these AR/VR specific hardware devices was described (see Table 7).

**Table 7**. Hardware Devices for AR/VR Jobs

| Rank | Hardware Devices (Percentage Count of AR/VR jobs) |
|---|---|
| 1 | Facebook Oculus Rift (33.05%) |
| 2 | HTC Vive (19.13%) |
| 3 | Microsoft Hololens (17.39%) |
| 4 | Samsung Gear (9.57%) |
| 5 | Varjo VR (7.83%) |

The most popular device identified was Oculus Rift, followed by HTC Vive and Microsoft Hololens. Due to the high initial investment cost in hardware for developing an AR/VR course, these findings could serve as a guideline to help institutions select the most appropriate AR/VR hardware. The desktop game development is primarily performed on two game engines: Unity and Unreal. Unity is a proprietary software developed by Unity Technologies, while Unreal is a commercial software developed by Epic Games. Based on our dataset, we observed that Unity (59.55%) was more popular than Unreal (40.45%). Overall, the current job listings demand more technical expertise in commercial software tools from corporations like Autodesk, Unity, etc.

Employee traits have also started to become more important for success in the workplace. First, as noted in Table 2, the Employee skill category was less valued than the Occupational attributes. Among the Employee skill category, time management was the most in-demand. Especially in the current business climate with millennials, time management skills are especially crucial. Employees with a focus on achieving deadlines in a timely manner are sought after in a fast-paced and dynamic workplace. The next ranked skill was motivation. Higher levels of motivation are typically associated with lifelong success in any profession. The next skill in the sorted list was attention to detail. It is regarded as an indicator of organized and attentive professionals who are paramount to the success of any firm. The next key soft skill was independence. An independent worker typically completes assigned activities with minimal supervision. These individuals are regarded as proactive and take the initiative towards ownership and completion of their assigned projects. Note that these employee attributes are valued for success in any job. These skills could be incorporated into the course offerings to adjust the curriculum appropriately.

Another important skill category deals with basic communication skills. The most frequent generic skills included keywords corresponding to responsibility and determination. Thus, taking initiatives and dealing with stressful situations were considered as success factors in a fast-paced work environment. Upon closer inspection, verbal, written and presentation skills were also valued for success in these job positions. This job entails understanding client requirements and typically modeling them through 3D software. Hence, various inputs from multiple team members need to be gathered and the employees are expected to analyze the aggregate results. Therefore, verbal and written skills are more in-demand. The presentation skills are also vital for AR/VR specialists since the modeling phase usually involves various meetings and brainstorming sessions with different teams.

Interpersonal skills were also valued for this job position. These skills typically involved cooperation and collaboration within a team environment. Personality and creativity attributes were also highly valued. Because, in various settings, the individual deliverables were assigned to each team member. Additionally, creativity is paramount in the art design field, which has many parallels with AR/VR. The analytical and research skill categories target individuals who were analytical thinkers and could use rational judgment to understand and resolve client requirements. The long-term growth of a company is contingent on such individuals with sound judgment. The creativity or innovation skillset pertains to the artistic side of the individual in tune with the arts design aspect of the job. Lastly, the analytical skillset coincides with the problem-solving mindset of the individual, which is necessary for the long-term growth of a firm. It is well known that all strategic problem-solvers benefit from a sound understanding of the business processes to increase the effectiveness of deployed solutions. Thus, the employees are supposed to have an essential knowledge of the holistic product lifecycle. This necessitates the understanding of product design, supply chains, and marketing channels to be successful professionals. Moreover, various routine administrative tasks are expected to be handled by AR/VR specialists.

## CONCLUSIONS AND FUTURE RESEARCH

This paper analyzed the skill requirements of AR/VR jobs in the US. The study employed content analysis to develop a ranking of the required skill categories. The findings found that occupational skills were most valued, technical in nature and placed a greater emphasis on UI/UX design, software design, asset design and graphics rendering skills. The job positions also assigned more importance to interpersonal and communication skills.

This research project established and ranked skills that are currently required for AR/VR positions. The results could be utilized in two different ways. First, the degree programs such as Computer Science and Fine Arts could repurpose their existing curriculum related to AR/VR. Hence, this paper could benefit the associated undergraduate and graduate degree programs. Second, the findings could be used by human resource departments to identify qualified AR/VR professionals. More specifically, the human resources department could design well-structured and consistent job descriptions to effectively manage their employees.

Future research could investigate the effects of job localization in some specific US states on different skill requirements. This would involve studying the job postings in a specific geographical region allowing local employers and universities to benefit from the relevant findings. An additional research direction could involve the analysis of the existing curricula of degree programs in the AR/VR domain in the US. In this way, the findings of the current study could be used in conjunction with the coursework of related degree programs. By identifying the existing skill gaps between academic programs and local industry needs, specific recommendations related to curriculum design could be made.

**REFERENCES**

Aasheim, C. L., Williams, S., Rutner, P., & Gardiner, A. (2015). Data Analytics vs. Data Science: A Study of Similarities and Differences in Undergraduate Programs Based on Course Descriptions. Journal of Information Systems Education, 26(2), 103-115.
Adi Robertson. (2021, July 7). Oculus Sales spike up in the lead up to Half-Life. Retrieved from https://www.theverge.com/2020/4/30/21241995/facebook-q1-2020-earnings-oculus-rift-quest-sales-half-life-alyx-pandemic
Akçayır, M. and Akçayır, G., 2017. Advantages and challenges associated with augmented reality for education: A systematic review of the literature. Educational Research Review, 20, pp.1-11.
Anton, E., Behne, A., & Teuteberg, F. (2020). The humans behind artificial intelligence – an operationalization of AI competencies.
Bacca Acosta, J.L., Baldiris Navarro, S.M., Fabregat Gesa, R. and Graf, S., 2014. Augmented reality trends in education: a systematic review of research and applications. Journal of Educational Technology and Society, 2014, vol. 17, núm. 4, p. 133-149.
Bacca Acosta, J.L., Baldiris Navarro, S.M., Fabregat Gesa, R. and Graf, S., 2014. Augmented reality trends in education: a systematic review of research and applications. Journal of Educational Technology and Society, 2014, vol. 17, núm. 4, p. 133-149.
Billinghurst, M., Clark, A. and Lee, G., 2015. A survey of augmented reality.
Chi, H.L., Kang, S.C. and Wang, X., 2013. Research trends and opportunities of augmented reality applications in architecture, engineering, and construction. Automation in construction, 33, pp.116-122.

De Swert, K., 2012. Calculating inter-coder reliability in media content analysis using Krippendorff's Alpha. Center for Politics and Communication, 15.

Debortoli, S., Müller, O., & vom Brocke, J. (2014). Comparing business intelligence and big data skills. Business & Information Systems Engineering, 6(5), 289-300.

Doshi, A., Smith, R.T., Thomas, B.H. and Bouras, C., 2017. Use of projector based augmented reality to improve manual spot-welding precision and accuracy for automotive manufacturing. The International Journal of Advanced Manufacturing Technology, 89(5-8), pp.1279-1293.

Facebook. (2021, July 7). AR/VR – New Dimensions of Connection. Retrieved from https://www.facebook.com/business/news/insights/future-ar-vr

Fominykh, M., Wild, F., Klamma, R., Billinghurst, M., Costiner, L.S., Karsakov, A., Mangina, E., Molka-Danielsen, J., Pollock, I., Preda, M. and Smolic, A., 2020. Model augmented reality curriculum. In Proceedings of the Working Group Reports on Innovation and Technology in Computer Science Education (pp. 131-149).

Fominykh, M., Wild, F., Klamma, R., Billinghurst, M., Costiner, L.S., Karsakov, A., Mangina, E., Molka-Danielsen, J., Pollock, I., Preda, M. and Smolic, A., 2020, June. Developing a model augmented reality curriculum. In Proceedings of the 2020 ACM Conference on Innovation and Technology in Computer Science Education (pp. 508-509).

Fominykh, Mikhail & Bilyatdinova, Anna & Koren, István & Jesionkowska, Joanna & Karsakov, Andrey & Khoroshavin, Aleksandr & Klamma, Ralf & Klimova, Alexandra & Molka-Danielsen, Judith & Rasool, Jazz & Smith, Carl & Wild, Fridolin. (2019). Existing Teaching Practices and Future Labour Market Needs in the Field of Augmented Reality. 10.13140/RG.2.2.33051.49443/2.

Gardiner, A., Aasheim, C., Rutner, P., & Williams, S. (2018). Skill requirements in big data: A content analysis of job advertisements. Journal of Computer Information Systems, 58(4), 374-384.

Goldiez, B., Livingston, M. A., Dawson, J., Brown, D., Hancock, P., Baillot, Y., & Julier, S. J. (2004, November 29–December 2). Advancing human-centered augmented reality research [Paper presentation]. 24th Annual Army Science Conference, Orlando, FL. https://apps.dtic.mil/sti/pdfs/ADA433480.pdf

Helin, K., Karjalainen, J., Bolierakis, S., Frangakis, N., Kemppi, P., Tedone, D. and Oliveira, D.M., 2017. Augmented Reality System for Space Station Maintenance Support. In European Association for Virtual Reality and Augmented Reality Conference, EuroVR-2017.

Hired. (2021, July 17). 2021 State of Software Engineers Retrieved from https://hired.com/state-of-software-engineers

Ibáñez, M.B. and Delgado-Kloos, C., 2018. Augmented reality for STEM learning: A systematic review. Computers & Education, 123, pp.109-123.

James Paine. (2021, July 17). 10 Brands Already Leveraging the Power of Augmented Reality. Retrieved from https://www.inc.com/james-paine/10-brands-already-leveraging-power-of-augmented-reality.html

Joseph Seb. (2021, July 7). How Ikea is using Augmented Reality. Retrieved from https://digiday.com/media/ikea-using-augmented-reality/

Lovaglio, P. G., Cesarini, M., Mercorio, F., & Mezzanzanica, M. (2018). Skills in demand for ICT and statistical occupations: Evidence from web-based job vacancies. Statistical Analysis and Data Mining: The ASA Data Science Journal, 11(2), 78-91.

Mourtzis, D., Zogopoulos, V. and Vlachou, E., 2017. Augmented reality application to support remote maintenance as a service in the robotics industry. Procedia Cirp, 63, pp.46-51.

Nee, A.Y., Ong, S.K., Chryssolouris, G. and Mourtzis, D., 2012. Augmented reality applications in design and manufacturing. CIRP annals, 61(2), pp.657-679.

Neuendorf, K. A. (2016). The Content Analysis Guidebook. Sage Publications: Thousand Oaks, CA.

Perry, T.S., 2020. AR/VR is this year's hot ticket for jobs: But growth in demand for blockchain developers stutters-[Careers]. IEEE Spectrum, 57(4), pp.19-19.

Robert Brown. (2021, July 7). Augmented Reality in the Future of Work. Retrieved from https://www.cognizant.com/futureofwork/article/augmenting-your-future-job-title-and-monetization-with-augmented-reality

Robert Half. (2021, July 7). The Future of Work – Adapting to Technical Change. Retrieved from https://www.roberthalf.com/research-and-insights/workplace-research/the-future-of-work

Saari, A., Rasul, M.S., Yasin, R.M., Rauf, R.A.A., Ashari, Z.H.M. and Pranita, D., 2021. Skills Sets for Workforce in the 4th Industrial Revolution: Expectation from Authorities and Industrial Players. Journal of Technical Education and Training, 13(2), pp.1-9.

Schmalstieg, D. and Hollerer, T., 2016. Augmented reality: principles and practice. Addison-Wesley Professional.

Thomas Alsop. (2021, July 7). Global Augmented Virtual Reality Market Size. Retrieved from https://www.statista.com/statistics/591181/global-augmented-virtual-reality-market-size/

Thornton, T., Ernst, J. V., & Clark, A. C. (2012). Augmented reality as a visual and spatial learning tool in technology education. Technology and Engineering Teacher, 71(8), 18–21.

Verma, A., Frank, P. and Lamsal, K., 2021. An exploratory study of skill requirements for social media positions: A content analysis of job advertisements. arXiv preprint arXiv:2106.11040.

Verma, A., Lamsal, K. and Verma, P., 2021. An investigation of skill requirements in artificial intelligence and machine learning job advertisements. Industry and Higher Education, p.0950422221990990.

Verma, A., Yurov, K. M., Lane, P. L., & Yurova, Y. V. (2019). An investigation of skill requirements for business and data analytics positions: A content analysis of job advertisements. Journal of Education for Business, 94(4), 243-250.

Wang, M., Callaghan, V., Bernhardt, J., White, K. and Peña-Rios, A., 2018. Augmented reality in education and training: pedagogical approaches and illustrative case studies. Journal of ambient intelligence and humanized computing, 9(5), pp.1391-1402.

Wang, M., Callaghan, V., Bernhardt, J., White, K. and Peña-Rios, A., 2018. Augmented reality in education and training: pedagogical approaches and illustrative case studies. Journal of ambient intelligence and humanized computing, 9(5), pp.1391-1402.

Wilson, H. J., Daugherty, P., & Bianzino, N. (2017). The jobs that artificial intelligence will create. MIT Sloan Management Review, 58(4), 14.